\documentclass{jpsj3}
\usepackage{color}
\usepackage{txfonts}
\usepackage[T1]{fontenc}
\usepackage{subfigure}

\title{Online Calibration Scheme for Training Restricted Boltzmann Machines with Quantum Annealing}

\author{Takeru Goto$^{1,2}$\thanks{takeru{\_}goto@jp.honda}, Masayuki Ohzeki$^{1,3,4}$}
\inst{$^1$Graduate School of Information Sciences, Tohoku University, Miyagi 980-8564, Japan \\
$^2$Innovative Research Excellence, Honda R\&D Co., Ltd., Tokyo 107-6238, Japan \\
$^3$Department of Physics, Tokyo Institute of Technology, Tokyo 152-8551, Japan \\
$^4$Sigma-i Co., Ltd., Tokyo 108-0075, Japan} %\\

\abst{We propose a scheme to calibrate the internal parameters of a quantum annealer to obtain well-approximated samples for training a restricted Boltzmann machine (RBM).
Empirically, samples from quantum annealers obey the Boltzmann distribution, making them suitable for RBM training. 
Quantum annealers utilize physical phenomena to generate a large number of samples in a short time. 
However, hardware imperfections make it challenging to obtain accurate samples. Existing research often estimates the inverse temperature for the compensation.
Our scheme efficiently utilizes samples for RBM training also to estimate internal parameters.
Furthermore, we consider additional parameters and demonstrate that they improve sample quality. We evaluate our approach by comparing the obtained samples with classical Gibbs sampling, which theoretically generates accurate samples. Our results indicate that our scheme demonstrates performance on par with Gibbs sampling.
In addition, the training results with our estimation scheme outperform those of the contrastive divergence algorithm, a standard training algorithm for RBM.
}

\begin{document}
\maketitle

\section{Introduction}
Deep neural networks have gained widespread attention due to their excellent practical performance.
Restricted Boltzmann machines (RBMs), as a subset of deep neural networks, are used for unsupervised and supervised tasks. Their application include facial recognition\cite{liu2014facial,he2013facial}, natural language understanding\cite{sarikaya2014application}, predictions of electricity loads and degradation of manufacturing facilities\cite{dedinec2016deep,wang2019deep}.

An RBM is trained by minimizing the Kullback-Leibler (KL) divergence between the model distribution and an empirical distribution composed of training data\cite{ackley1985learning}.
Training an RBM requires statistical mean values of samples generated by the RBM. Although Gibbs sampling is commonly used to achieve this, it incurs a high computational cost due to its complexity and the long equilibration time required for samples to converge to an accurate distribution. The contrastive divergence (CD) algorithm is frequently applied to address this\cite{hinton2002training}. CD estimates the gradient of the RBM’s parameters using Gibbs sampling with limited sampling steps, reducing computational costs. While not guaranteed to yield fully accurate results, CD provides a computationally efficient approximation compared to full Gibbs sampling.

As an alternative to the schemes mentioned above, the quantum annealing (QA)\cite{kadowaki1998quantum} approach has attracted attention for its ability to sample from a physical device directly. QA controls the strength of the transverse field in an adiabatic process, wherein the initial state of superposition changes to low-energy states of the final user-designed Hamiltonian. 
QA has been mainly studied as a solution for optimization where it works by searching for the ground state of the Ising model\cite{lucas2014ising}, as in traffic flow optimization\cite{neukart2017traffic,hussain2020optimal,inoue2021traffic}, 
 finance\cite{rosenberg2016solving, orus2019forecasting, venturelli2019reverse}, logistics \cite{feld2019hybrid,ding2021implementation}, manufacturing\cite{venturelli2015quantum, Yonaga2022, Haba2022}, preprocessing in material experiments\cite{Tanaka2023}, marketing \cite{nishimura2019item}, and decoding problems \cite{IdeMaximumLikelihoodChannel2020, Arai2021code}.
Its potential for solving optimization problems with inequality constraints has been enhanced \cite{yonaga2020solving}, particularly in cases where the direct formulation is challenging\cite{Koshikawa2021}. 
A comparative study of quantum annealer was performed for benchmark tests to solve optimization problems \cite{Oshiyama2022}. 
The quantum effect on the case with multiple optimal solutions has also been discussed \cite{Yamamoto2020, maruyama2021graph}. 
As the environmental effect cannot be avoided, the quantum annealer is sometimes regarded as a simulator for quantum many-body dynamics \cite{Bando2020, Bando2021, King2022}. 
Furthermore, applications of quantum annealing as an optimization algorithm in machine learning have also been reported \cite{neven2012qboost,khoshaman2018quantum,o2018nonnegative,Kumar2018,Arai2021}. In addition to these, the use of quantum annealers as samplers for Boltzmann distribution has also been researched. \cite{Sato2021,Amin2018,Urushibata2022,hasegawa2023kernel}.

While quantum annealing effectively solves optimization problems, its application in Boltzmann machine learning has also been established.
Empirical evidence supports the assertion that the samples obtained from the quantum annealer closely approximate the Boltzmann distribution.
Accordingly, several studies have demonstrated its application to the training of Boltzmann machines \cite{adachi2015application,dixit2021training,benedetti2016estimation,xu2021adaptive}. Some of them\cite{adachi2015application, dixit2021training} have reported results superior to those of the CD algorithm and faster than those of the Gibbs sampling.

Obtaining appropriate samples without calibration is difficult because of hardware imperfections, such as chain breaks and working temperature. These imperfections are quantitatively represented by internal parameters, such as the inverse temperature $\beta_{\rm{eff}}$. To address this, $\beta_{\rm{eff}}$ of the generated samples is often estimated as a way to compensate for the parameters of couplings and fields that are mapped to the quantum annealer. Adachi et al.\cite{adachi2015application} determined the best fit $\beta_{\rm{eff}}$ that minimizes the error of interaction term of visible and hidden nodes before model training. This scheme can not follow the temporal variation of $\beta_{\rm{eff}}$. As a simple approach, Dixit et al.\cite{dixit2021training} trained models with multiple values of $\beta_{\rm{eff}}$ and adopted the best result. However, such an approach wastes significant computational resources to train models that would remain unused. To handle these problems, several studies\cite{benedetti2016estimation,xu2021adaptive} have estimated the parameter in the model training process. Benedetti et al.\cite{benedetti2016estimation} derived a relationship among the difference of probabilities, the energies, and the $\beta_{\rm{eff}}$ of two samples. They then used the relationship with linear regression on multiple samples to approximate $\beta_{\rm{eff}}$. This approach requires post-processing during each iteration, which also impacts training time. The estimation scheme of Xu et al.\cite{xu2021adaptive} infers $\beta_{\rm{eff}}$ by maximizing the likelihood. 
However, the computational cost incurred by this approach is also impractical, as it must calculate the average value over all configurations of the RBM. Although this is not specified clearly, we assume that Gibbs sampling is employed for the approximation, further increasing the training time.

Our method uses samples from a quantum annealer as training data and applies the CD algorithm to estimate internal parameters. The primary objective is to learn the Boltzmann machine parameters. The CD algorithm for inferring internal parameters minimizes the total training time because it does not require significant computational resources. The samples from the quantum annealer are also used to train the model for updating the Boltzmann machine parameters, similar to training with Gibbs sampling. As learning progresses, the internal parameters are well-calibrated. By the end of the learning process, since the samples closely follow the Boltzmann distribution, our scheme is expected to achieve performance very close to that of Gibbs sampling. In addition, we introduce three patterns of assumptions with regards to internal parameters: (i) one inverse temperature as in the methods mentioned above, (ii) three inverse temperatures for weights, biases of visible units, and biases of hidden units, (iii) one inverse temperature for weights, and a coefficient for each bias.

The remaining part of this paper consists of the following sections. The next section presents the update rule for estimated internal parameters and a detailed algorithm for inferring them during training. In the subsequent section, we showcase the experiments' results using the D-Wave 2000Q to generate samples. The implemented RBM has 32 visible nodes and 8 hidden nodes. Although it is relatively small, it suffices to validate the proposed scheme. The first half of this section focuses on investigating how the proposed estimation scheme improves sample quality without training. We compare the proposed scheme with conventional Gibbs sampling using histograms and the KL divergence of the samples. We observe that the increased number of internal parameters enhances the sampling quality. In the latter part of this section, we perform a training test to demonstrate the results of our scheme compared to classic schemes. Similar to previous research, our scheme is more accurate than the CD algorithm. The number of internal parameters exhibits the same effect as in the previous experiment.

\section{Methods}
\subsection{Theory of RBM}
An RBM consists of a visible layer and a hidden layer denoted as $v=\{v_1,v_2, \ldots, v_n\}$ and $h=\{h_1,h_2, \ldots, h_m\}$, respectively. Each layer has stochastic binary variables, and the probability follows a Boltzmann distribution as
\begin{equation}
    P({v},{h}|\theta)=\frac{\exp{\left(-E({v},{h}|\theta)\right)}}{Z},\label{eq:prob}
\end{equation}
where $\theta$ is the vector of the RBM's parameters, $E$ is the energy function expressed as
\begin{equation}
    E({v},{h}|\theta)=-\sum_{i,j}w_{ij}v_{i}h_{j} - \sum_{i}b_{i}v_{i} - \sum_{j}c_{j}h_{j},\label{eq:energy}
\end{equation}
and $Z$ is the partition function $Z=\sum_{{v},{h}}P(v,h|\theta)$.
The RBM is trained by minimizing the KL divergence, expressed as
\begin{equation}
    D_\mathrm{KL}(Q_D||P)=\sum_{v\in{\chi_v}}Q_D(v)\log\frac{Q_D(v)}{P(v|\theta)}, \label{eq:kl}
\end{equation}
where ${\chi_v}$ is a set of all configurations for the visible layer of the RBM, $Q_D$ is the empirical distribution of the training data, and $P(v|\theta)$ is a marginal distribution of Eq. (\ref{eq:prob}).
Minimizing the KL divergence is equivalent to maximizing the log-likelihood as
\begin{equation}
    \mathcal{L}(\theta)=\sum_{v\in{D}}Q_D(v)\log{P(v|\theta)}, \label{eq:loglikelihood}
\end{equation}
where $D$ is the set of training data, and $N_D$ is its size.
As a result, we obtain the gradient of $\mathcal{L}$ with respect to $\theta$ as
\begin{equation}
    \frac{d\mathcal{L}}{d\theta}=\left<\frac{\partial{E}}{\partial\theta}\right>_D-\left<\frac{\partial{E}}{\partial\theta}\right>_{\theta}, 
\end{equation}
where $\left<f(v,h)\right>_D$ and $\left<f(v,h)\right>_\theta$ are averages over the training dataset and all configurations, respectively. These expectations are defined as follows:
\begin{equation}
\begin{split}
    \left<f(v,h)\right>_D&=\sum_{v\in{D}}Q_D(v)\sum_{h\in{\chi_h}}P(h|v,\theta)f(x,v), \\
    \left<f(v,h)\right>_\theta&=\sum_{v\in{\chi_v}}P(v|\theta)\sum_{h\in{\chi_h}}P(h|v,\theta)f(x,v) \\
    &=\sum_{v,h\in{\chi}}P(v,h|\theta)f(x,v).
\end{split}    
\end{equation}
Here, $\chi_h$ denotes a set of all configurations for the hidden layer and $\chi$ is a Cartesian product ${\chi_v}\times{\chi_h}$.
Each parameter can be updated by using the gradient method according to the following rules:
\begin{equation}
\begin{split}
    &\delta{w_{ij}}=\eta\left(\left<{v_i}{h_j}\right>_D-\left<{v_i}{h_j}\right>_{\theta}\right),\\
    &\delta{b_{i}}=\eta\left(\left<v_i\right>_D-\left<v_i\right>_{\theta}\right),\\
    &\delta{c_{j}}=\eta\left(\left<h_j\right>_D-\left<h_j\right>_{\theta}\right). \label{eq:upd_rbm}
\end{split}
\end{equation}
Calculating the $\left<\ldots\right>_{\theta}$ is difficult because the number of configurations is $2^n$. Instead, Gibbs sampling is applied to estimate it. Furthermore, to reduce the calculation cost, the CD algorithm\cite{hinton2002training} generates samples evolved from the training data instead of calculating $\left<\ldots\right>_{\theta}$. Although these results are not exact, the gradient direction is sufficiently approximated. Consequently, the model can be trained over a single evolution.

\subsection{Training RBM with quantum annealer}
\begin{figure}[t]
\centering
\includegraphics[width=0.9\linewidth]{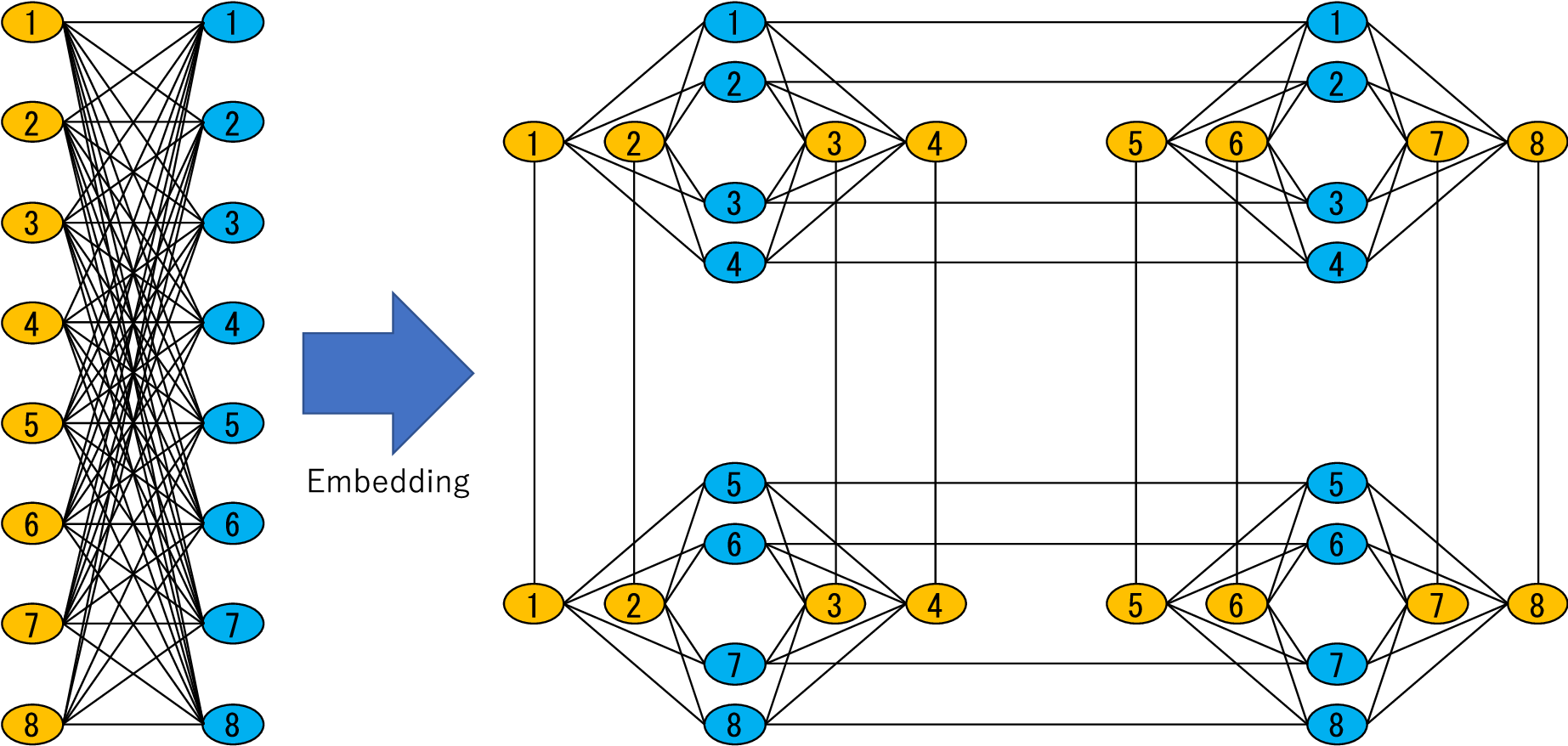}
\caption{An RBM with a dimension of $8\times8$, and its embedding to a Chimera graph-based hardware. (Color online)}
\label{fig:embedding}
\end{figure}

As an alternative to Gibbs sampling, quantum annealing can be used to collect samples for the estimation of $\left<\ldots\right>_\theta$ in Eq. (\ref{eq:upd_rbm}). Due to the hardware structure, the number of couplings among qubits is limited. Hence, each node in the RBM is embedded in multiple qubits in the quantum annealer. Figure \ref{fig:embedding} illustrates a sample embedding with D-Wave 2000Q, which has a Chimera graph structure\cite{adachi2015application}. The details of the hardware mapping of an RBM to a quantum annealer and the scheme of constructing QUBO matrix are provided in the literature \cite{adachi2015application}.

When a QUBO matrix is determined by Eq. (\ref{eq:energy}), prior studies\cite{adachi2015application} have assumed that samples from a quantum annealer are
obtained as
\begin{equation}
    P({v},{h}|\theta,\beta_{\rm{eff}})=\frac{\exp{\left(-\beta_{\rm{eff}}E({v},{h}|\theta)\right)}}{Z_{\beta_{\rm{eff}}}},\label{eq:beta_eff}
\end{equation}
where $\beta_{\rm{eff}}$ is an internal parameter of the quantum annealer, and the partition function $Z_{\beta_{\rm{eff}}}$ is $\sum_{{v},{h}}P(v,h|\theta,\beta_{\rm{eff}})$. Samples following Eq. (\ref{eq:prob}) as opposed to Eq. (\ref{eq:beta_eff}) are needed to estimate $\left<\ldots\right>_\theta$. Once $\beta_{\rm{eff}}$ is estimated, appropriate samples are generated from the quantum annealer by constructing a QUBO matrix as $E({v},{h}|\theta)/\beta_{\rm{eff}}$, which adjust the actual energy landscape by negating the internal parameter. Thus, the obtained samples will follow the distribution in Eq. (\ref{eq:prob}), allowing the term $\left<\ldots\right>_\theta$ in Eq. (\ref{eq:upd_rbm}) to be approximated with the returned samples. The details of the hardware mapping of an RBM to a quantum annealer and constructing the QUBO matrix scheme are provided in the literature \cite{adachi2015application}.

As mentioned, most studies have adopted the inverse temperature as an internal parameter. However, as the quantum annealer's hardware has many unknown aspects, whether this correction is sufficient is unclear. To address this issue, we formulate the difference between the sample distribution generated by quantum annealing and the original RBM using KL divergence. We then minimize this difference to accommodate other patterns of internal parameters. The method proposed below uses the CD algorithm, which requires less calculation, similar to training an RBM.

\subsection{Inferring Internal Parameters}
\begin{figure*}[!t]
\centering
\includegraphics[width=1.0\linewidth]{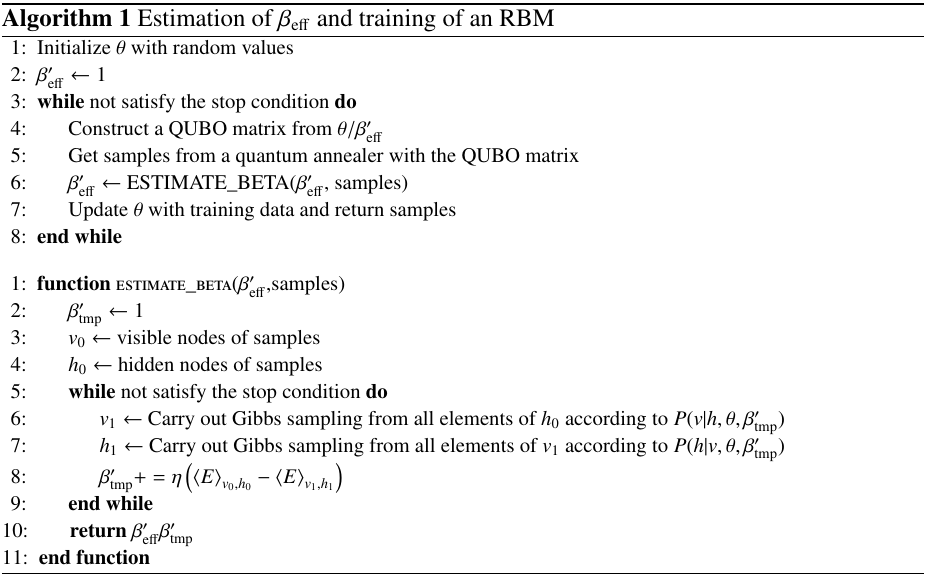}
\caption{The proposed algorithm. $\left<\ldots\right>_\theta$ in Eq. (\ref{eq:update_beta_eff}) is replaced by an ensemble average of the samples evolved by Gibbs sampling from the return samples with only one or a few iterations owing to the CD algorithm. This differs from full Gibbs Sampling, which involves many iterations to reach equilibrium. The stop conditions for loops were specified as 1500 times for training and three times for estimating $\beta_{\rm{eff}}$.}
\label{fig:algorithm}
\end{figure*}

We propose a scheme to infer the internal parameters $\beta$ of a quantum annealer during the training of an RBM. The scheme uses the return samples from the quantum annealer as training data for estimation, wherein $\theta$ is fixed during estimation. Although our conclusions are closely aligned with the findings in the literature \cite{xu2021adaptive}, we derive the formulation similar to Eqs. from (\ref{eq:kl}) to (\ref{eq:upd_rbm}). This enables explicit handling of samples that include both visible and hidden layer bits obtained from the quantum annealer. Our objective is to estimate the value of $\beta$ that minimizes the KL divergence between $P(v,h|\theta,\beta)$ and the empirical distribution of the samples $Q_S(v,h)$ defined as
\begin{equation}
    D_\mathrm{KL}(Q_S||P)=\sum_{v,h\in{\chi}}Q_S(v,h)\log\frac{Q_S(v,h)}{P(v,h|\theta,\beta)}. \label{eq:kl_beta}
\end{equation}
Here, $Q_S(v,h)=N_{v,h}/N_S$, where $N_{v,h}$ represents the number of samples with the configuration $v$ and $h$ among the returned samples, and $N_S$ is the total number of the samples. Similar to Eq. (\ref{eq:loglikelihood}), this is equivalent to maximizing the log-likelihood:
\begin{equation}
    \mathcal{L}(\beta) = \sum_{v,h\in{S}}Q_S(v,h)\log{P(v,h|\theta,\beta)}, \label{eq:loglikelihood_beta}
\end{equation}
where $S$ is the set of samples. Unlike the case of training an RBM, training data, which are samples from a quantum annealer, include hidden layer bits. Thus, the probability distribution in Eq. (\ref{eq:loglikelihood_beta}) does not represent a marginal distribution of a visible layer. As a result, the gradient of Eq. (\ref{eq:loglikelihood_beta}) with respect to $\beta$ is derived as
\begin{equation}
    \frac{d\mathcal{L}}{d\beta}=\left<\frac{\partial{E}}{\partial\beta}\right>_S-\left<\frac{\partial{E}}{\partial\beta}\right>_{\theta}, \label{eq:dldb}
\end{equation}
where $\left<f(v,h)\right>_S$ is the average over the returned samples, defined as follows:
\begin{equation}
\begin{split}
    \left<f(v,h)\right>_S&=\sum_{v,h\in{S}}Q_S(v,h)f(x,v).
\end{split}
\end{equation}
If $\beta_{\rm{eff}}$ in Eq. (\ref{eq:beta_eff}) is assumed as an internal parameter, 
the updating rule for $\beta_{\rm{eff}}$ is derived using Eq. (\ref{eq:dldb}):
\begin{equation}
    \delta\beta_{\rm{eff}}=\eta\left(\left<E\right>_S-\left<E\right>_\theta\right). \label{eq:update_beta_eff}
\end{equation}

The overall algorithm combining the estimation of $\beta_{\rm{eff}}$ with training is shown in Fig. \ref{fig:algorithm}.
We define \(\beta'_{\rm{eff}}\) as the estimated value, in contrast to the true \(\beta_{\rm{eff}}\). The compensated samples follow the distribution as 
\begin{equation}
\begin{split}
    P({v},{h}|\theta/\beta'_{\rm{eff}},\beta_{\rm{eff}})&=\frac{\exp{\left(-\beta_{\rm{eff}}E({v},{h}|\theta/\beta'_{\rm{eff}})\right)}}{Z_{\beta_{\rm{eff}}}}\label{eq:beta_eff'} \\
    &=\frac{\exp{\left(-\beta_{\rm{eff}}/\beta'_{\rm{eff}}E({v},{h}|\theta)\right)}} {Z_{\beta_{\rm{eff}}}} \\
    &=P({v},{h}|\theta,\beta_{\rm{tmp}}),
\end{split}
\end{equation}
where $\beta_{\rm{tmp}}$ is $\beta_{\rm{eff}}/\beta'_{\rm{eff}}$. If the estimated value of $\beta_{\rm{tmp}}$, denoted as $\beta'_{\rm{tmp}}$, is appropriately obtained, $\beta'_{\rm{tmp}}\beta'_{\rm{eff}}$ in line 10 of Fig. \ref{fig:algorithm} is closer to $\beta_{\rm{eff}}$ than $\beta'_{\rm{eff}}$.
The CD algorithm can be used to update $\beta'_{\rm{tmp}}$ similar to the standard training procedure for RBM. The typical CD algorithm needs three-step Gibbs sampling from the visible layers' bits of the training data. However, in our cases, the samples from QA already have hidden layers' bits. Thus, we approximate Eq. (\ref{eq:update_beta_eff}) with two-step Gibbs sampling as the line 6 to 8 in Fig. \ref{fig:algorithm}.
To estimate the effective temperature in the early stages of training, it is necessary to acquire many samples in advance. However, these samples are not effective for RBM training due to the lack of proper calibration. Therefore, we estimate the effective temperature simultaneously with the training process. Although the update amount of the effective temperature in a single iteration is small, it is gradually estimated over many iterations, similar to the training process. This approach is justified qualitatively because the accuracy of the sample distribution becomes more critical in the later stages of training than in the early stages.

The sample distribution is strongly correlated with the original Boltzmann distribution. Especially, the estimation of the inverse temperature $\beta_{\rm{eff}}$
enables the learning of RBM with the returned samples\cite{adachi2015application}. However, the mechanism of the internal errors in quantum annealers is not sufficiently understood. Due to the possibility that the distribution of samples has a different energy landscape, we try to improve it by introducing two patterns of additional calibration parameters and distributions:
\begin{equation}
\begin{split}
        P({v},{h}|\theta,\beta_{vh},\beta_v,\beta_h)=\frac{\exp{\left(\beta_{vh}\sum_{i,j}w_{ij}v_{i}h_{j}+\beta_{v}\sum_{i}b_{i}v_{i}+\beta_{h}\sum_{j}c_{j}h_{j}\right)}}{Z_{\beta_{vh,v,h}}},\label{eq:beta_vh_v_h}
\end{split}
\end{equation}
and
\begin{equation}
\begin{split}
    &P({v},{h}|\theta,\beta_{vh},\beta_{v1},\ldots,\beta_{vn},\beta_{h1},\ldots,\beta_{hm})= \\ &\qquad\frac{\exp{\left(\beta_{vh}\sum_{i,j}w_{ij}v_{i}h_{j}+\sum_{i}\beta_{vi}b_{i}v_{i}+\sum_{j}\beta_{hj}c_{j}h_{j}\right)}}{Z_{\beta_{vh,vi,hi}}}.\label{eq:beta_vh_vi_hi}
\end{split}
\end{equation}
We derive the update rule from Eq. (\ref{eq:dldb}) similar to Eq. (\ref{eq:update_beta_eff}) for each internal parameter of Eq. (\ref{eq:beta_vh_v_h}) as
\begin{equation}
\begin{split}
    &\delta{\beta_{vh}}=\eta\left(\left<\sum_{i,j}w_{ij}v_{i}h_{j}\right>_S-\left<\sum_{i,j}w_{ij}v_{i}h_{j}\right>_\theta\right),\\
    &\delta{\beta_{v}}=\eta\left(\left<\sum_{i}b_{i}v_{i}\right>_S-\left<\sum_{i}b_{i}v_{i}\right>_\theta\right),\\
    &\delta{\beta_{h}}=\eta\left(\left<\sum_{j}c_{j}h_{j}\right>_S-\left<\sum_{j}c_{j}h_{j}\right>_\theta\right),\label{eq:upd_each_beta1}\\ 
\end{split}
\end{equation}
and for Eq. (\ref{eq:beta_vh_v_h}) as
\begin{equation}
\begin{split}
    &\delta{\beta_{vi}}=\eta\left(\left<b_{i}v_{i}\right>_S-\left<b_{i}v_{i}\right>_\theta\right),\\
    &\delta{\beta_{hj}}=\eta\left(\left<c_{j}h_{j}\right>_S-\left<c_{j}h_{j}\right>_\theta\right).\label{eq:upd_each_beta2}\\ 
\end{split}
\end{equation}
These patterns of internal parameters are hereinafter referred to as one-parameter, three-parameter, and one-and-all-bias, respectively. The estimation of many internal parameters requires a large number of samples. Because the proposed method is an online estimation scheme, an increase in internal parameters leads to an increase in estimation time. If we introduce a calibration parameter for each cross term $v_ih_j$, the increase in parameters is the product of $n$ and $m$, which may significantly increase the estimation time. We therefore focus on the above three patterns of internal parameters.

\section{Experiments}
\subsection{Simulation}

\begin{figure*}[t]
\centering
\subfigure[Only biases have errors]{
\includegraphics[width=7.5cm]{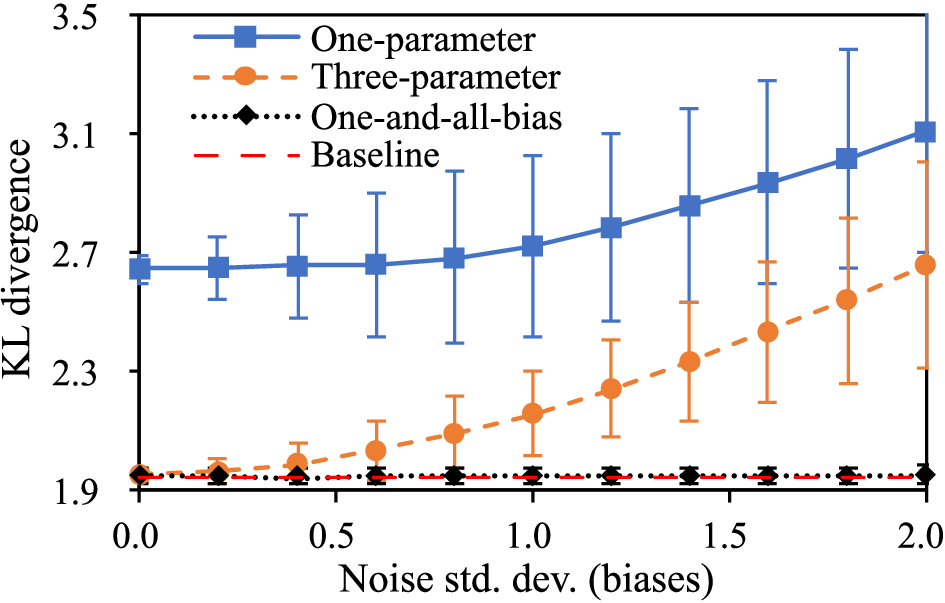}
}
\subfigure[Both weights and biases have errors]{
\includegraphics[width=7.5cm]{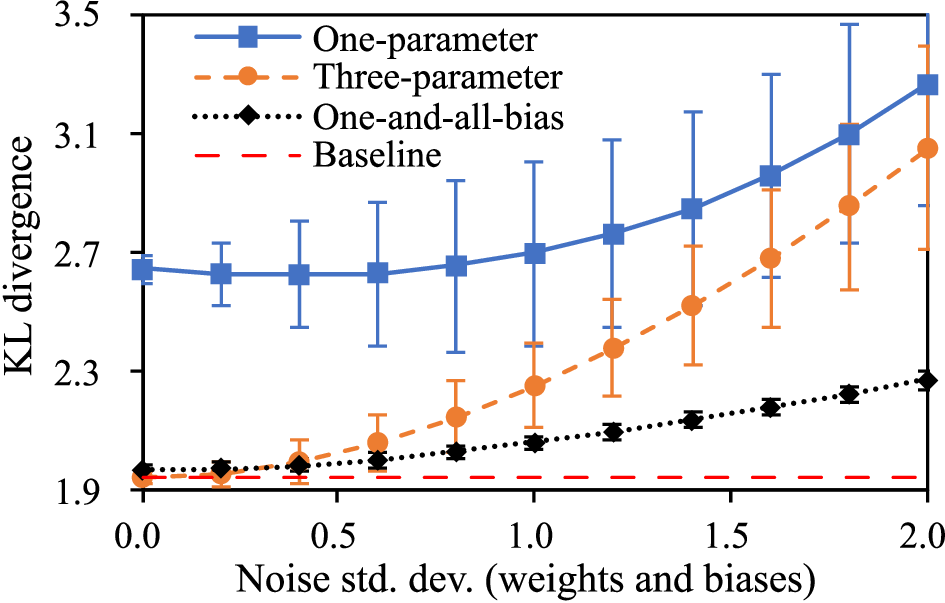}
}
\caption{The result of the KL divergence between the original Boltzmann distribution and 1,000,000 samples generated under a simulated noise environment. The baseline represents the result of samples generated in a no-noise environment. In (a), all $\beta_{{\rm{err}}_{wij}}$ are set to 6.8. In (b), each $\beta_{{\rm{err}}_{wij}}$ follows $\mathcal{N}(6.8, \sigma)$. In common, each $\beta_{{\rm{err}}_{bi}}$ and $\beta_{{\rm{err}}_{cj}}$ follows $\mathcal{N}(7.0, \sigma)$ and $\mathcal{N}(4.5, \sigma)$, respectively. These specific values are based on the result of the actual hardware test described below. The statistical mean and standard deviation were calculated over 10 simulations. (Color online)}
\label{fig:simulation}
\end{figure*}

To validate the internal parameters' estimation part can improve the quality of samples, we performed simulations using an RBM pre-trained by the CD algorithm. The RBM consisted of 32 visible nodes and 8 hidden nodes, trained on a coarse-grained MNIST dataset, similar to that used in prior studies\cite{adachi2015application}. The simulation environment introduced errors mimicking actual hardware characteristics, generating samples based on the energy function as 
\begin{equation}
\begin{split}
    &E({v},{h}|\theta, \beta_{\rm{err}})=-\sum_{i,j}\beta_{{\rm{err}}_{wij}}w_{ij}v_{i}h_{j} - \sum_{i}\beta_{{\rm{err}}_{bi}}b_{i}v_{i} - \sum_{j}\beta_{{\rm{err}}_{cj}}c_{j}h_{j},\label{eq:error}
\end{split}
\end{equation}
where $\beta_{\rm{err}}$ is a vector representing the hardware errors. We measured the KL divergence between the original RBM distribution and the samples obtained from the simulation environment calibrated by the estimated internal parameters $\beta$. 

Figure \ref{fig:simulation} shows the simulation test result. In (a) of Fig. \ref{fig:simulation}, all $\beta_{{\rm{err}}_{wij}}$ are set to the same value. Each $\beta_{{\rm{err}}_{bi}}$ and $\beta_{{\rm{err}}_{cj}}$ follows a normal distribution. In this case, the one-and-all-bias has enough theoretical calibration parameters. The result also indicates this, as there is no degradation of the KL divergence as the error strength of each bias increases. In (b) of Fig. \ref{fig:simulation}, each $\beta_{{\rm{err}}_{wij}}$ also follow a normal distribution. Even though the calibration parameters are insufficient to fully compensate for the hardware noise, increasing the parameters significantly improves the quality of the samples.

\subsection{Actual Hardware}

\begin{table*}[t]
\centering
\caption{KL divergence between empirical distributions of samples and the original RBM.}
\label{tab:kl}
\begin{tabular}{lllll}
\hline
Number of samples & Gibbs sampling & One-parameter & Three-parameter & One-and-all-bias \\
\hline
100000 & 3.49 & 3.88 & 3.70 & 3.64 \\
1000000 & 1.94 & 2.46 & 2.31 & 2.13\\
\hline
\end{tabular}
\end{table*}

\begin{figure*}[t]
\centering
\includegraphics[width=17cm]{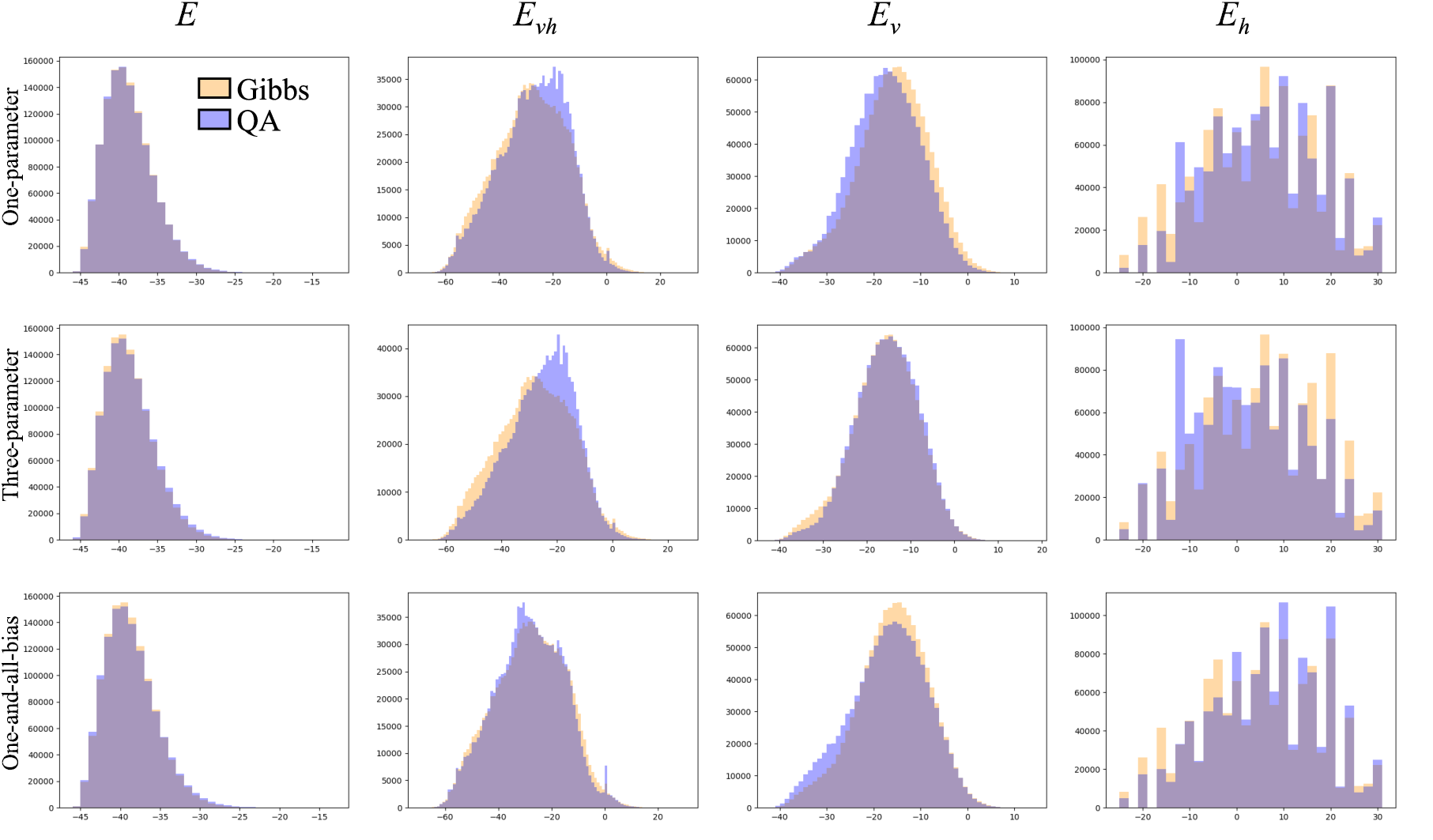}
\caption{Histograms of samples' energy. To investigate the effect of internal parameters, we separate the total energy $E$ into $E_{vh}=-\sum{w_{ij}v_ih_j}$, $E_{v}=-\sum{b_iv_i}$ and $E_{h}=-\sum{c_jh_j}$. (Color online)}
\label{fig:histogram}
\end{figure*}

\begin{figure*}[t]
\centering
\vspace{5mm}
\includegraphics[width=16cm]{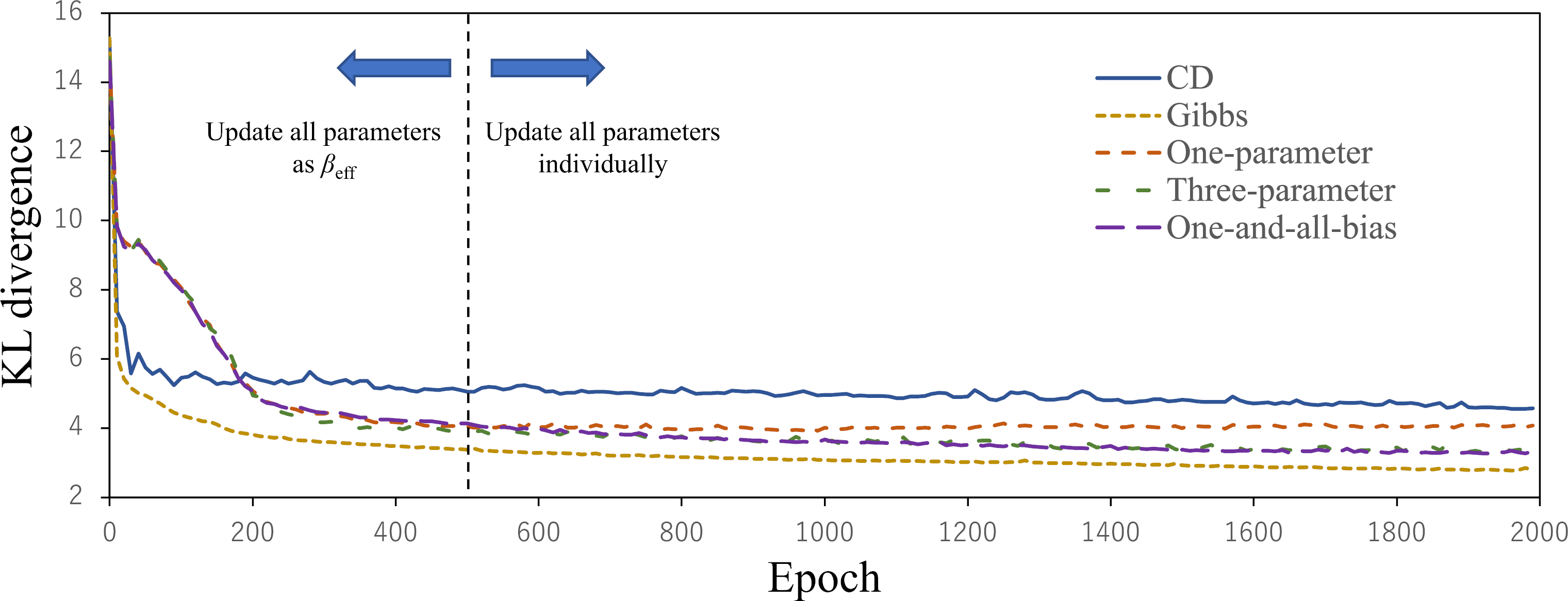}
\vspace{-5mm}
\caption{Transition of KL divergence between the empirical distribution of the training data and the trained RBM of each training scheme. The minimum KL divergence values for CD, Gibbs, One-parameter, Three-parameter, and One-and-all-bias are 4.56, 2.78, 3.92, 3.30, and 3.26, respectively. (Color online)}\label{fig:learning}
\end{figure*}

\begin{figure*}[t]
\centering
\includegraphics[width=16cm]{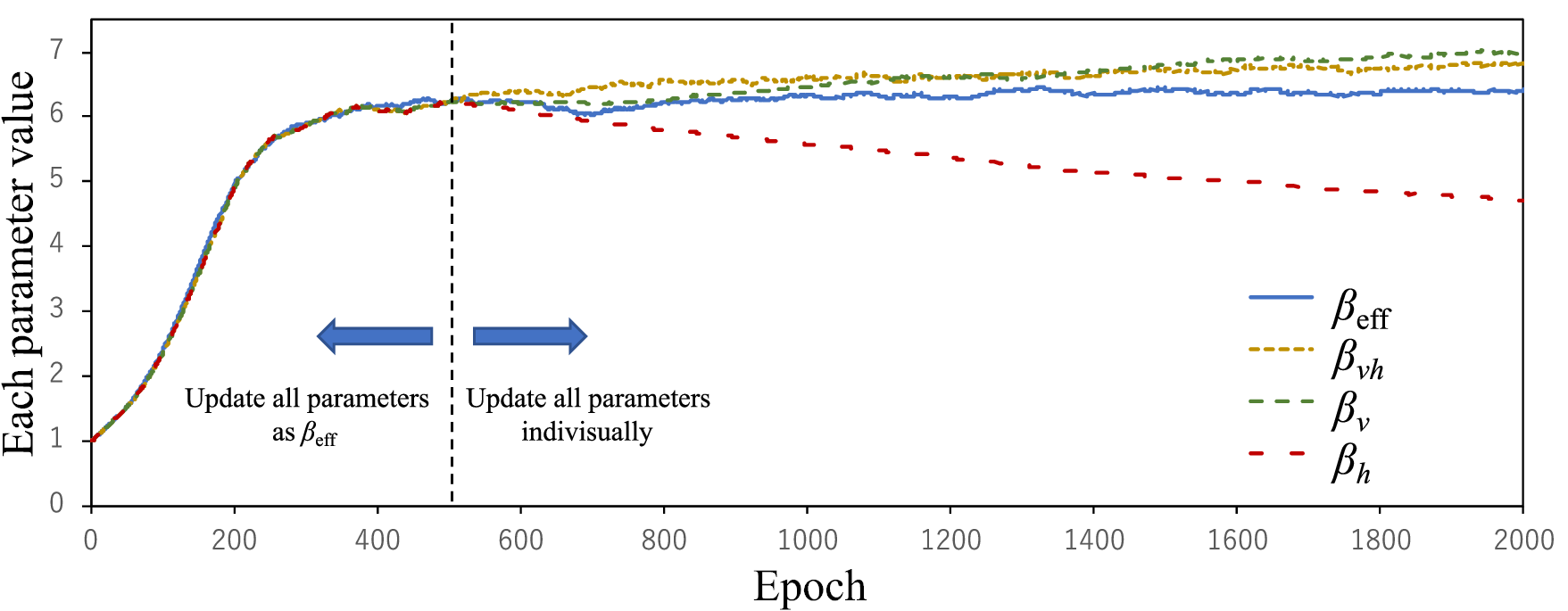}
\vspace{-5mm}
\caption{Transition of the estimated internal parameters. We only illustrate the parameters of the one-parameter and three-parameter cases, as there are too many parameters in the one-and-all-bias case. (Color online)}
\label{fig:beta}
\end{figure*}

The same RBM was embedded in 12 different locations on the D-Wave 2000Q, avoiding faulty qubits to gather samples efficiently. We compared samples generated by Gibbs sampling and quantum annealing with those calibrated by the proposed scheme. Table \ref{tab:kl} shows the results of KL divergence. The increase in the size of samples decreased KL divergence because the empirical distributions approach the original distribution. In theory, Gibbs sampling can accurately obtain samples following the Boltzmann distribution. Therefore, results close to those obtained by Gibbs sampling can be considered more accurate. We found that increased internal parameters improved sample quality, as indicated by the simulation result. To capture the characteristics of the samples returned from the quantum annealer, Fig. \ref{fig:histogram} presents histograms of the samples' energies. Although one parameter can adjust the total energy $E$, the samples' distributions of $E_{vh}$ and $E_v$ exhibit differences between Gibbs sampling and quantum annealing. $E_v$ of the three-parameter case and $E_{vh}$ of one-and-all-bias case seem to fit well with Gibbs sampling. Quantum annealers have not only temperature differences but also qubit imperfections. We consider that many internal parameters can be calibrated to improve sample quality, even for actual hardware.

Next, we conducted RBM training with the simultaneous estimation of internal parameters. The RBM size and training data were identical to the previous experiment. The number of times for updating $\beta$ in each epoch was set to five. The increase in the number of internal parameters enlarges the estimation time. Therefore, all parameters were updated following the same approach as $\beta_{\rm{eff}}$ using Eq. (\ref{eq:update_beta_eff}) for up to 500 epochs. Then, each parameter was updated with the corresponding update rule. We compare our scheme's results with the conventional CD algorithm and Gibbs sampling. Figure \ref{fig:learning} presents the transition of KL divergence between the empirical distribution of training data and each trained RBM. The results obtained by the proposed algorithm are above those of the CD algorithm and below those of Gibbs sampling. The one-and-all-bias case exhibited the best results of the proposed three-parameter configurations, correlating with the previous experimental results. The convergence speed of our scheme slows down just after starting. However, training rapidly progresses between 100 and 200 epochs. We present the estimation results of internal parameters in Fig. \ref{fig:beta}, indicating that the process accelerates training. As mentioned, all parameters were almost equivalent under 500 epochs as the same update rule was applied. This also indicates that $\beta_{\rm{eff}}$ converges at approximately 500 epochs, leading to the stagnation of KL divergence in the one-parameter case, as illustrated in Fig. \ref{fig:learning}. On the other hand, the variation of other parameters continued after 500 epochs, and the KL divergence results of two other parametric configurations exhibited continuous improvement. Therefore, these results are better than those of the one-parameter case, revealing that the increase in internal parameters is effective for training. However, whereas $\beta_{\rm{eff}}$ converged early, $\beta_h$ did not especially converge. This indicates that the estimation of many internal parameters requires a large number of samples.

\section{Discussion}
We propose a novel scheme for estimating the internal parameters $\beta$ of a quantum annealer during the training of an RBM. Our scheme is developed by minimizing the KL divergence between the empirical distribution of samples obtained from the quantum annealer and a Boltzmann distribution constructed using the RBM and $\beta$. Accordingly, the CD algorithm is adopted to infer the parameters. Samples are utilized for both training and estimation simultaneously, which reduces training time compared to conventional schemes. In addition, the proposed algorithm can calibrate multiple internal parameters. Although the inverse temperature is generally estimated as an internal parameter, using additional internal parameters potentially enhances the sample distribution's accuracy.

Through both simulation and real hardware calibration results, we confirmed that an increased number of internal parameters reduces KL divergence. The hardware results do not reach the baseline, particularly for the one-and-all-bias with the most internal parameters. In simulations, we found that the calibrated samples can achieve the same quality as the baseline if the parameters are theoretically sufficient. That indicates that the hardware has non-negligible noise on the weights among qubits or some errors that might distort the expected Boltzmann distribution. Understanding the characteristics of the real hardware’s errors may help introduce more suitable internal parameters.

Comparing the training results, we see that the proposed method surpasses the conventional CD algorithm, as indicated by our results. However, it is important to note that estimating numerous parameters using our method requires a significant amount of time. Additional techniques appear to be required, such as initiating an update of all parameters in the middle of training and increasing the update rate just after starting. Indeed, in our experiments, the inverse temperature was estimated before 500 epochs, and all other parameters initiated updating from the estimated inverse temperature. Indeed, more trials are needed to fully understand the quantitative effects and the potential of our scheme for the actual hardware.

Furthermore, the impact on training time may vary depending on factors such as the number of internal parameters, the size of the RBM, the amount of training data, and other hyperparameters. We are interested in examining these relationships for practical use. In addition, developing an effective scheme to estimate many internal parameters is important. Furthermore, Amin et al.\cite{amin2018quantum} proposed the quantum Boltzmann machine (QBM) as an extension of RBM. Noisy quantum devices, such as the quantum annealer, may be used to train QBMs. A fine-tuning of internal parameters in the devices may also improve performance, as indicated by our experiments. Accordingly, we must consider a suitable calibration scheme that can evolve from the proposed method for QBMs.

\bibliographystyle{jpsj}
\bibliography{71374}

\end{document}